# [Research Paper] Formalizing Interruptible Algorithms for Human *over*-the-loop Analytics

Austin Graham
School of Computer Science
University of Oklahoma
Norman, Oklahoma 73071
austin.p.graham-1@ou.edu

Yan Liang
School of Computer Science
University of Oklahoma
Norman, Oklahoma 73071
yliang@ou.edu

Le Gruenwald
School of Computer Science
University of Oklahoma
Norman, Oklahoma 73071
ggruenwald@ou.edu

Christan Grant
School of Computer Science
University of Oklahoma
Norman, Oklahoma 73071
cgrant@ou.edu

*Abstract*—Traditional data mining algorithms are exceptional at seeing patterns in data that humans cannot, but are often confused by details that are obvious to the organic eye. Algorithms that include humans "in-the-loop" have proved beneficial for accuracy by allowing a user to provide direction in these situations, but the slowness of human interactions causes execution times to increase exponentially. Thus, we seek to formalize frameworks that include humans "over-the-loop", giving the user an option to intervene when they deem it necessary while not having user feedback be an execution requirement. With this strategy, we hope to increase the accuracy of solutions with minimal losses in execution time. This paper describes our vision of this strategy and associated problems.

## I. Introduction

Data mining algorithms have been unimaginably successful at discovering patterns and groupings in data humans cannot naturally perceive. Following this success, one problem has become apparent: although computing has allowed the automation of discovery tasks, analysts are still required to oversee and configure parameters to achieve the most accurate solution. Without the data context and expertise possessed by the analyst, algorithms must be executed multiple times to experiment with various configurations. As data sets grow exponentially in size and algorithms becoming increasingly complex, such tasks are becoming temporally expensive.

For this reason, it would be useful if an analyst could dynamically adapt a learning method to changing contexts. For example, an effective data streaming clustering must be able to easily adjust for rapidly changing data sets. Points are added as data flows in, and removed as it flows out; re-running an algorithm on this type of data would produce inconsistent and out-of-date models.

A more common situation is during a long mining process, an analyst might notice something about the current state that is inconsistent or incorrect. Instead of stopping the process and restarting the algorithm and potentially repeating useful work, the analyst could instead adjust the parameters during the current execution and thus take advantage of the correct work done to that point.

This type of problem has been explored from various directions. It has been the goal of many analysts to limit the number of times an algorithm must be executed and to boost the accuracy by giving the algorithm context via the help of a subject matter expert. Although these experiments have been generally successful, involving humans in the algorithmic process has seen much controversy; some subsets of analysts believe subject matter experts *help* algorithms by allowing intervention in outlier cases, others believe systems involving them can *hinder* discovery. Miettinen [1] cites the original purpose of data mining: to discover patterns and information within data that have not or could not have been discovered by a human. In the case of human interactivity with these algorithms, she argues "the user [will] steer the algorithm towards the kind of results that she a priori considered useful and interesting, and never find the kind of results she did not expect to find." Indeed, if done incorrectly, human intervention could heavily bias the results of discovery towards information he or she already knows, defeating the purpose of the process. Similarly, Orseau and Armstrong [2] apply interactivity with reinforcement learning, fearing human intervention would bias the policy towards what the analyst already believes to be optimal. If this is found to be the result, then again we have defeated the purpose of learning.

In addition, most real-world data sets exist in dimensions outside the visual spectrum. For a human to interact with an algorithmic process, it is necessary they *comprehend* the process. While systems such as one proposed by Bond et al. [3] find potential solutions to projecting high-dimensional information to the visual spectrum, it is still difficult to comprehend the changes of thousands or millions of data points at a time. It is more appropriate to convey the shape and relationships between the data via the learning method to understand what is being discovered and what potential errors might require human intervention.

It is thus our goal to define a class of algorithms that we deem "interruptible". These algorithm place the human "over-the-loop", that is, humans can oversee the operations and progress of the algorithm and can make changes as they are necessary with minimal increases in execution time, but if no changes need to be made then the algorithm will execute as normal. As with the previous examples, it is a concern of ours that allowing this interaction would yield only suspected results; we seek to define interactions that will only help alleviate execution time and limit the number of times an analyst must run an algorithm to achieve acceptable results. To accomplish this goal, we present *Human-Over-the-Loop-Analytics* (HOLA), an architecture for communicating attributes of an arbitrary learning method such that an analyst can easily pinpoint and address errors as they happen, without restarting the algorithm and wasting resources.

## II. BACKGROUND

Human-interactive algorithms have been proposed with one of three architectures: interactive, anytime, and interruptible. A discussion of other proposed interruptible algorithms is saved until later, but it is important to note the distinction between interruptible and interactive and anytime.

Interactive algorithms place the human "in-the-loop"; a human-dependent step is included within the description of algorithm execution. In other words, the human is an active part of the algorithm. Given the iterative nature of many data mining tasks, interactive algorithms tend to place human interaction at discrete, end-of-iteration times. One such iterative-style clustering has found success with this architecture; algorithms such as the one proposed by Awasthi et al. [4] that allows users to split or merge clusters was found to produce well-defined classes of data points, or another like Vikram and Dasgupta [5], which allows the user to specify constraints of sub clusters, found higher data likelihood than the non-interactive counterparts. Other such algorithms, like that of Lad and Parikh [6] or Biswas and Jacobs [7] define confusion states where the algorithm will ask for human input, and were found to have clearer data classes. At the extreme algorithms like the one proposed by Amershi et al. [8] are entirely user driven, that is, the algorithm only takes actions on user input and were also found to define much clearer data classes.

These algorithms have found success in improving model quality at a cost of detrimental setbacks in execution time. The human-dependent steps require the algorithm stop and wait for the user input, thus if the user is not available the algorithm will not continue. Therefore, availability is sacrificed for model quality; a human must be ready to annotate for the duration of execution to finish the task.

Anytime algorithms were developed to alleviate the need for users to be present and available machine execution. These algorithms give the user control over the "execution time" versus "model quality" trade off — the algorithm executes as needed, providing a solution at every point during the execution. This allows the user to develop the model to an acceptable point of accuracy before terminating the process. Ueno et al. [9] applies this process to nearest neighbor classification and Vlachos et al. [10] with KMeans clustering, both examples eventually reaching an anytime state where further iterations simply refine the model and will continue do so forever or until instructed to stop and deliver a solution.

While anytime algorithms give analysts more control over execution time versus model quality, they still do not address the issue of re-running potentially several times. If a parameter or weight needs to be changes the program must be restarted. Although the concept of anytime algorithms implies a longer amount of time spent executing will give finer results, these results will be refined to the potentially inaccurate parameters given. Analysts are still required to experiment with configurations to achieve maximum model quality.

Unlike interactive or anytime algorithms, interruptible algorithms address both the time and accuracy issue. The allowance for the user to change the state mid-execution saves the experimentation time of re-running the algorithm in its entirety with new parameters. Orseau and Armstrong [2] provides an example where a robot learns to either carry boxes off of a truck into a warehouse, or to stack boxes within the warehouse based off of a set of parameters. However, if it rains it should not go outside to the truck, thus human interruption in learning would be a necessity. It is this situation which inspired this research: we seek to assist computational tasks handle situations which cannot be understood by the learning method.

It is essential that this new process does not increase the hardness of the learning problem; below we present requirements a change must have to be considered interruptible in an algorithm.

## III. THE INTERRUPTIBLE ALGORITHM

Identifying an interruptible algorithm requires insight: proposing an algorithm as interruptible where a change is more complex than that of the algorithm itself could potentially yield execution times consistently greater than that of re-running the algorithm. For this reason, we propose methods to measure the interruptibility of an algorithm and how the interruptibility may scale.

We define an interruption in an Interruptible Algorithm as changes made in the *core data set*, *data model*, or *hyper parameters* that result in an execution complexity of less than that of the algorithm itself. In mathematical terms, if a change $C$ is made to attribute $U$ in algorithm $A$ and $O(A) > O(C)$, then we say that algorithm $A$ is interruptible in attribute $U$.

### A. Interruptiblilty Index

For comparison, we define an "interruptibility index" for an algorithm $A$, change $C$, and attribute $U$:

$$I_{A,U} = O(A)/O(C) \quad (1)$$

We can then see that if $I_{A,U} > 1$, then the complexity of the algorithm is larger than that of the change, thus the change is negligible compared to the execution of the algorithm. As the index grows higher, we can see that $A$ becomes increasingly interruptible in attribute $U$. If $I_{A,U} \leq 1$ then the complexity of the change is equal to or greater than that of the algorithm, and it can be concluded that re-running the algorithm would be a more efficient use of resources. A similar conclusion can be made that as the index approaches zero, $A$ is said to be decreasingly interruptible in attribute $U$.

This index will often include attributes of the inputs to the algorithms that can be used to determine the expense of the interrupt. This measurement would allow analysts to measure the expense of the interrupt against the perceived quality of the current model and other possible interrupts. For example, two interrupts $A$ and $B$ may improve the quality of a working model within an acceptable range, but one may see that the target of interrupt $A$ has an interruptibility index that is quadratic with hyper parameter $U_1$, and the target of interrupt $B$ has an interruptibility index that is linear with hyper parameter $U_1$. Thus, interrupt $A$ is more expensive than interrupt $B$, as we can see that increasing $U_1$ will quadratically increase the runtime of the algorithm. The following section illustrates these properties with concrete examples.

### B. Example: Interrupting KMeans

Take an example of a data change in an implementation of the KMeans clustering algorithm. It is known that the complexity of this algorithm is $O(nkI\theta)$ where $n$ is the number of points, $k$ is the number of clusters, $I$ is the number of iterations the algorithm executes before termination, and $\theta$ is the dimensionality of all points MacQueen et al. [11]. Consider some data point $q$ changed mid-iteration. We find that the adjustment of the cluster to which $q$ belongs can be done in $O(\theta)$ time. Thus for this data change in $q$ for the KMeans algorithm, by (1) we see that the interruptibility index is $I_{KMeans,q} = O(nkI\theta)/O(\theta) = nkI$.

The interruptibility index is dependent on the number of iterations $I$, number of clusters $k$, and total number of points $n$. If $\theta$ were to be present in this formula, then we can conclude that the change in a data point would increase the complexity of the KMeans algorithm linearly. The lack of this term implies that this change is constant relative to the total algorithm.

Alternately, consider a change $a$ to the hyper parameter $k$. This change would involve the creation of a new cluster definition (constant time) with the movement of a single point ($O(\theta)$) into this new cluster so it may be considered. For simplicity, we assume this point is chosen at random. Thus, the interruptibility index could be calculated the same as the aforementioned data change: $I_{KMeans,a} = O(nkI\theta)/O(\theta) = nkI$. The index is the same, but the $k$ hyper parameter is included in this formula, thus the complexity of KMeans increases linearly with increases in $k$. In this manner, we see that change in point $q$ is less expensive than change in $k$, and we have a notion of the overall affect of the runtime from this change. Additionally, an analyst could compute the numerical value of this index to get concrete measurements for interruptibility and compare interrupts more effectively.

### C. Summary

In this section, we defined the interruptibility index and explored its implications in the context of an interruptible algorithm. However, the changes to the algorithm are not the only items that need to be considered when implementing an interruptible framework. The next section discusses other issues related to interruptibility and associated problems that will need to be solved to achieve a reasonable solution.

## IV. RESEARCH QUESTIONS

When implementing an interruptible algorithm, several considerations must be made to determine its feasibility. There are many algorithms that have data structured in a way that is inherently difficult for humans to understand, or execute to quickly for reasonable human interaction. In the following sections, we discuss these issues and explore potential solutions.

### A. Data Set Applicability

For an interruptible algorithm to be feasible, the underlying data set must be appropriate for analyst interaction. Human reaction time is very slow compared to that of a computer: while the human brain can operate in milliseconds, computers operate in nano seconds; once a analyst has noticed a change that has been made and is able to process and manipulate the data, the computer will have advanced several steps in its operations.

The core problem we are trying to solve involves humans interacting with computational tasks during their operations, thus for an interrutible algorithm to be effective we must guarantee that the user will be able to perceive alterations as they happen. With a data set that allows an algorithm to execute in a time interval the human perceives as instant, the wait-time associated with experimentation becomes negligible; no subsequent process can improve upon this speed. It can then be concluded that interruptible algorithms should apply to only large data sets with processing time of hours or days such that the current state of the data can be communicated, altered, and re-communicated by the human in time to be applied before the algorithm has completed execution.

### B. Data Visualization

For a human to be involved with a computational task, it is necessary they comprehend the information being communicated to them. Machine learning algorithms and models in particular are less intuitive; a user can easily manipulate inputs and parameters, but the resulting structure of the model in many cases can not be easily understood. For example, the effects of a change in the size of a hidden layer in an Artificial Neural Network can be estimated, but a guess-and-check process is needed to tune the model effectively. Clustering models have optimal tolerance parameters dictating what falls in each cluster, all of which vary between data sets. Such parameters are difficult to surmise; trial and error becomes essential in determining the best settings for a particular data set.

Effective model parameters for a particular data set are subjective, thus difficult to compute *a priori*. The analyst must try various combinations, or run expensive processes to determine the optimal settings to solve the given problem. This is time consuming, and in many cases the analyst can determine what should be adjusted mid-run of an algorithm.

In this manner, it is the responsibility of the visualization system to make the model *interpretable*, or understandable to humans for its appropriate uses Doshi-Velez and Kim [12]. When a model is made interpretable, users can extrapolate information and take appropriate actions quicker, aiding particularly in time and accuracy sensitive fields such as health care and national defense. In these situations, a user does not have the time nor the resources to discern the meaning of every data point and make the proper changes; they need to be provided with a fluid interface that conveys enough information to determine quality and accuracy of the model. Systems proposed by Khan et al. [13] and Rahman et al. [14] are examples of this interface style. In both cases, the focus of the visualization is not on the data points, but rather the operations and trends performed on them.

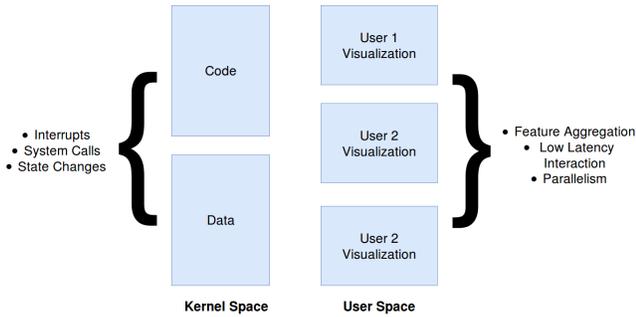

Fig. 1. Overview of HOLA architecture.

However successful, these systems are not without their flaws. Both systems are very specialized to a single type of system; Khan et al. [13] focuses on relational databases and Rahman et al. [14] focuses on simple data sampling. Goals of interruptibility include not only understanding the data, but the model as well thus an effective visualization system would need to aggregate this information into one cohesive form for any generic model.

## V. System Architecture

An initial implementation of the HOLA system outlines three key features: (1) Visualize model state information to user, (2) Allow a user to per scribe a change to an algorithm component, and (3) Handle interrupts intermittently through execution, concurrent to the visualization. Figure 1 shows an overview of the architecture consisting of two operating system inspired components: a kernel space and a user space. It is our goal this system be *collaborative* and aim to have multiple users experimenting on a single data set. The kernel acts as an independent system that broadcasts information to multiple users, and thus should be able to serve, weight, and handle multiple user interrupts. To accomplish this goal, both the user and kernel space must be completely independent; several components must be defined to allow for interaction between the two. These components are defined in the following sections.

### A. Kernel Space

The kernel space contains the learning algorithm implementation and data storage mechanisms. This section has four main components: an interrupt priority queue, the algorithm implementation, a data storage repository, and a state renderer.

The interrupt priority queue will accept interrupts in the form of system calls that have been sanitized from user input in the user space. The prioritization scheme is subject to customization by a user, but will have a default scheme that is the subject of future work. Table V lists the possible system calls for the HOLA kernel. These functions outline the original two goals of interruptibility: allow for user changes in data and changes in the data model, while additionally allowing for control over flow speed and location for easier analysis of execution. Function for the data model include update, remove, or add, allowing for more dynamic data points during execution. Learning model function include inspect and hupdate to pull relevant data information and change a hyper parameter respectively. Finally, an analyst can alter the current visualized state with cspeed, rewind, fforward, and pause.

The algorithm implementation is a custom version of a learning algorithm that has been decorated with "critical sections" that define where it is safe to make changes in the data or data model. Listing 1 illustrates the main function should be decorated with the interruptible handler that takes as parameters all changeable components of the algorithm. In this example, we have a simple KMeans implementation in the Python programming language. The changeable components of KMeans are the data points `data_points` and the hyper parameter `K`.

**Listing 1** Sample KMeans Interruptible Implementation
```
@hola(data_points, K)
def find_centers(data_points, K):
    oldmu = random.sample(X, K)
    mu = random.sample(X, K)
    while not has_converged(mu, oldmu):
        oldmu = mu
        hola.interrupt()
        clusters = cluster_points(X, mu)
        mu = reeval_centers(oldmu, clusters)
    return(mu, clusters)
```

The inline `hola.interrupt` function defines the critical sections where it is safe for the algorithm to make changes. Initially, this function is blocking: the algorithm will stop executing to handle the interrupts. However, it is a point of interest to have this be non-blocking and is saved for future research. In this KMeans implementation, we wish to handle changes before a new iteration so the changes can be taken into account sooner.

The data storage repository can be a database management system of the users choice. For security reasons, it is the responsibility of the kernel space to transform this information into a renderable form.

### B. User Space

The user space queries the kernel for visualization information, and handles user requests. Therefore, this section has two main components: the visualization engine, and the input interpreter.

The visualization engine asks the kernel for the newly transformed data and performs the rendering onto a window.

| System Call | Description |
|---|---|
| **Data** | |
| update(record r) | Update a data record |
| remove(record r) | Remove a data record |
| add(record r) | Add a data record |
| **Data Model** | |
| hupdate(parameter p, model m) | Update hyper parameter p for model m |
| inspect(model m) | Query the model for visualization |
| **Control** | |
| cspeed(speed s) | Alter speed of visualization stream |
| rewind(location l) | Rewind to visualization frame |
| fforward(location l) | Fast forward to visualization frame |
| pause() | Pause algorithm execution |
| open() | Open a connection with the kernel |
| close() | close a connection with the kernel |

TABLE I
THE HOLA SYSTEM CALLS.

It is essential that this process be adaptive enough to keep up with rapidly changing data so users are seeing current, relevant states on which to make changes. For greater interactivity, various forms of interfaces should be explored: web pages, mobile applications, virtual reality, and other such platforms can make the process more familiar, lessening the learning curve. In this way, the user can not only click on a screen or select buttons, but make changes through hand slicing motions or swiping to make changes more intuitive. With this design, there is the potential to re-use interaction styles between models to expedite the information extraction and improvement for any model.

The input interpreter is attached to the visualization engine and is responsible for accepting the user input and transforming it into the relevant system calls. Keeping with the KMeans example, a user might be using a virtual reality simulation engine and slice a cluster in half, effectively changing the $K$ hyper parameter. Such an action would be encoded as an hupdate command and sent to the interrupt priority queue in the kernel. Or the user might put one hand up and move the other to the side in a sliding motion, attempting to slow down the visualization. These motions would translate into the pause and rewind or fforward commands (depending on the direction of the sliding). In addition, we have included an open and close command representing a user joining and exiting a session.

## VI. CONCLUSION AND FUTURE WORK

In this paper we presented the HOLA architecture for interruptible algorithms, and methods for the analysis and action on data to alleviate pains of running and re-running algorithms. Although there would be slight increases in execution because of the needed overhead for the interrupts, this slow down is magnitudes smaller than the latter choice.

When implementing the theoretical framework described in this paper, we expect several research and design questions to arise. Some of these questions include the visualization of the data and the model, the prioritization of interrupts, and the quantifying the effectiveness of interruptibility.

An effective visualization system would to continuously be update the data and the model state to the user in such a way that the user can interpret and make appropriate changes within a span of time such that the interrupt is still relevant. Projects like Khan et al. [13], Rahman et al. [14], and Bond et al. [3] have made successful attempts at visualizing high dimensional data and complex processes in a digestable manner, however these projects operate on individual data points. With data sets approaching millions of points that are changing rapidly, these methods may become cumbersome for the users. Instead, we could propose a fluid-like interface that "flows" with the patterns discovered in the data, using viscosity and color to communicate model state and performance. Further analysis is saved for future research.

It is likely that an analyst would have priority of some changes over others during execution. For example, experiments being run on streaming data would want to prioritize the add, remove, or update system calls over others while experiments simulating a grid search for optimal parameters might prioritize the hupdate command over any model changes. The question of what is an effective default prioritization scheme remains open. This scheme must not allow for outdated interrupts to take precedent over more relevant ones, so we must implement this functionality into the default prioritization.

The purpose of an interruptible system to increase the comprehend ability of learning to a user, thus it is necessary that we conduct user studies as to the effectiveness not only to the results, but the effectiveness in usability of learning systems and ease of use. A simple strategy would be to obtain a comparable system for data mining and discovery, and allow multiple analysts to use both system for the same/similar problems. Thus, we can estimate the effectiveness of this new approach.

For increased collaboration, a git-like set of system calls that allow multiple users to experiment on the same data set concurrently could be added. This would allow analysts to work with the same visualization or each analysts could maintain local copies (branches) that keep changes to be merged into a master branch later. However, the kernel space-user space architecture must be altered: if a user branches the data, where are local changes kept? When does the system rebase a local branch with a master branch such that a user always has the most updated data while still reflecting their own changes? How are the interrupts made in a local branch reflected when merged into the master branch? These questions remain open, and are saved for a future research project.

Providing data analysis tools in this style will streamline the mining process and take pressure of the analyst to carefully pick parameters as to not waste hours or days. By enabling the analyst to point an algorithm in the correct direction, experiments will result in more accurate and adaptable models.

ACKNOWLEDGMENT

The authors would like to thank Stephen Smart and Nicholas Molloy for their early contributions to this research and the continued support.